\documentclass[11pt,a4paper]{article}
\usepackage{graphicx, amsmath}
\newcommand{\bo}{\boldsymbol}
\newcommand{\ff}[1]{\frac{1}{#1}}

\newcommand{\tj}[6]{\left(\begin{array}{ccc}#1 & #2 & #3\\#4 & #5 & #6\end{array}\right)}

\newcommand{\mean}[3]{\left<#1\left|#2\right|#3\right>}
\newcommand{\ee}[1]{(-)^{#1}}

\newcommand{\lc}{\left<}
\newcommand{\rc}{\right>}
\newcommand{\lr}{\left|}

\newcommand{\lb}{\left(}
\newcommand{\rb}{\right)}

\begin{document}

\begin{center}
{\Large RPA CORRELATIONS AND NUCLEAR DENSITIES IN RELATIVISTIC MEAN FIELD APPROACH }

\bigskip \bigskip

\small

N. VAN GIAI$^{1,2}$, H.Z. LIANG$^{1,2,3}$, J. MENG$^{3}$

\bigskip

\footnotesize

{\em $^1$Institut de Physique Nucl\'eaire, CNRS, UMR8608, F-91406 Orsay, France, E-mail: nguyen@ipno.in2p3.fr\\
$^2$Universit\'e Paris-Sud, F-91406 Orsay, France\\
$^3$School of Physics, Peking University, 100871 Beijing, China}

\bigskip

\small

(Dated: \today ) 

\end{center}

\bigskip

\footnotesize

{\em Abstract.\/} The relativistic mean field approach (RMF) is
well known for describing accurately binding energies and nucleon
distributions in atomic nuclei throughout the nuclear chart. The
random phase approximation (RPA) built on top of the RMF is also a
good framework for the study of nuclear excitations. Here, we
examine the consequences of long range correlations brought about
by the RPA on the neutron and proton densities as given by the RMF
approach.


\bigskip


\normalsize

\section{INTRODUCTION \label{sect1} }
The self-consistent mean field approach has become a powerful
theory for understanding the structure of atomic nuclei at a
microscopic level. The building blocks are structureless nucleons
interacting via two-body effective interactions. At lowest order
of the theory, correlations are neglected and the nucleus is
described as a system of $A$ particles moving independently in a
mean field created self-consistently by the two-body interactions.
This mean field is determined by applying a variational principle
to the total energy.

The above picture is known as the static Hartree-Fock (HF) theory.
A natural extension of it is the time-dependent Hartree-Fock
(TDHF) approach, whose linearized version coincides with the
well-known random phase approximation (RPA)\cite{Lane1960}. In
this way, long-range correlations of the particle-hole type are
introduced. They will bring corrections to various properties
predicted at the HF level, such as total energies, single-particle
spectra and occupation probabilities, densities, etc... In this
work, we want to focus on the effects of RPA long range
correlations on the nuclear densities. This question has been
studied several decades ago in the framework of HF-RPA
calculations done with the Gogny effective
interaction\cite{Gogny1979}. More recently Dupuis et
al.\cite{Dupuis2006} have re-examined the influence of RPA
correlations on nuclear densities and potentials, using again the
Gogny interaction.

However, besides the  non-relativistic HF and RPA framework there
is the relativistic counterpart represented by the relativistic
mean field (RMF) approach\cite{Serot1986,Ring1996,Meng2006} and
the relativistic RPA (RRPA) built upon the RMF\cite{Nicsic2002,
Vretenar2005}. The RMF approach consists in treating a
relativistic meson-nucleon effective Lagrangian at the level of
the Hartree approximation (no exchange) and no-sea approximation
(i.e., the Dirac sea is considered empty). Effective Lagrangians
with density-dependent meson-nucleon couplings have been
determined in order to give an excellent description of nuclear
ground states in RMF\cite{DDME2,Meng2006} as well as collective
excitations in RMF-RRPA\cite{Vretenar2005}. It is thus worthwhile
to examine the influence of RRPA correlations on the ground state
properties and to evaluate how much the RMF predictions will be
affected. This is what we will do in the rest of this paper,
concentrating the discussion on the nuclear densities.

\section{THEORY \label{sect2}}

\subsection{REMINDER OF RRPA \label{sub2.2}}

The starting point is an effective Lagrangian where the
nucleon-nucleon interaction is mediated by the exchange of
effective mesons (scalar-isoscalar $\sigma$, vector-isoscalar
$\omega$, vector-isovector $\rho$) and photons. The meson-nucleon
couplings are assumed to be density dependent. In general, the
parameters are adjusted so as to give a good description of
nuclear ground states throughout the periodic chart in the RMF
approach, i.e., in static Hartree plus no-sea approximation.
Having solved the RMF problem, one can build the complete set of
single-particle states consisting of the occupied states in the
Fermi sea, $h\equiv(e_h,l_h,j_h,m_h,\tau_h)$, and unoccupied
states $p\equiv(e_p,l_p,j_p,m_p,\tau_p)$, where the indices
$\tau_h(\tau_p)$ distinguish between neutrons and protons. Among
the latter states we must distinguish between unoccupied states in
the positive energy sector and those in the negative energy
sector. We will use the notation $\bar p$ and $\tilde p$ to
specify them if needed. The occurrence of $\tilde p$ states in
RRPA is a consequence of the no-sea treatment of the mean
field\cite{Ring2001}.

The RRPA formalism is well-known and we need only to recall the
main steps to make the notations clear. We denote by $b^\dag_p,
b^\dag_h$ the creation operators for states $p,h$. The creation
and annihilation operators of particle-hole states coupled to an
angular momentum $(JM)$ are:
\begin{eqnarray}
    c^\dag_{ph}(JM) &=& \sum_{m_pm_h}\ee{j_p-m_p}\hat J \tj{j_p}{j_h}{J}{m_p}{-m_h}{-M}b^\dag_{j_pm_p} b_{j_hm_h},\nonumber\\
    c_{ph}(J-M) &=& \sum_{m_pm_h}\ee{j_p-m_p}\hat J \tj{j_p}{j_h}{J}{m_p}{-m_h}{M}b^\dag_{j_hm_h} b_{j_pm_p},
\label{eq1}
\end{eqnarray}
where $\hat J=\sqrt{(2J+1)}$. In RPA, excited states $\vert\nu
JM\rangle$ are created by the action of  excitation operators
$Q_\nu^\dag(JM)$ on the correlated RPA ground state $\vert
RPA\rangle$. The excitation operators $Q_\nu^\dag(JM)$ are linear
combinations of particle-hole operators:
\begin{eqnarray}
    Q_\nu^\dag(JM) &=& \sum_{ph} X^{(\nu J)}_{ph}c^\dag_{ph}(JM)
        +\sum_{ph} Y^{(\nu J)}_{ph}c_{ph}(J-M)~,
\label{eq2}
\end{eqnarray}
where the $X$ and $Y$ amplitudes are determined by the RPA equations\cite{Ring1980}.

An important condition of the theory is that the correlated ground
state must be a vacuum for all RPA excitations:
\begin{equation}
    Q_\nu(JM)\lr \mbox{RPA}\rc=0~, \qquad \mbox{for all } \nu JM~.\\
\label{eq3}
\end{equation}

Then, it can be shown \cite{Rowe} that the RPA ground state can be
built explicitly from the HF ground state $\vert HF \rangle$:
\begin{equation}
    \lr \mbox{RPA}\rc = N_0 e^S \lr \mbox{HF}\rc~,
\label{eq4}
\end{equation}
where
\begin{equation}
    S=\ff2\sum_{JM}\sum_{php'h'}M_{php'h'}c^\dag_{ph}(JM)c^\dag_{p'h'}(J-M)~,
\label{eq5}
\end{equation}
and $N_0$ is a normalization coefficient. The coefficients
$M_{php'h'}$ are symmetric with respect to the $ph$ and $p'h'$
indices, $M_{php'h'}=M_{p'h'ph}$.  Note that, in all above
equations the summations over $p$ or $p'$ run on both unoccupied
positive energy (Fermi) $\bar p$ states and negative energy
(Dirac) $\tilde p$ states, as mentioned before.

An important relation between the $X$, $Y$ and $M$ coefficients
can be derived if one uses the condition (\ref{eq3}) that the RPA
ground state is a vacuum for the $Q_\nu(JM)$ operators. This
relation reads:
\begin{equation}\label{eq6}
    \sum_{p'h'}M_{php'h'}X^{(\nu J)}_{p'h'} = -Y^{(\nu J)}_{ph}~,
\end{equation}
and it will be used in the next subsection.

\subsection{ONE-BODY DENSITY MATRIX OF RPA GROUND STATE \label{sub2.1}}
Our main task in this work is to construct the one-body density
matrix of the correlated ground state,
$\rho_{\alpha\beta}=\mean{\mbox{RPA}}{b^\dag_\beta
b_\alpha}{\mbox{RPA}}$ and to compare it with that of the
uncorrelated HF state,
$\rho^{(0)}_{\alpha\beta}=\mean{\mbox{HF}}{b^\dag_\beta
b_\alpha}{\mbox{HF}}$.

We first observe that all quantities $\rho_{ph}$ and $\rho_{hp}$
must vanish because operator products of the type $b^\dag_h b_p$
and $b^\dag_p b_h$ are just linear combinations of $Q^\dag$ and
$Q$, and neither $\mean{\mbox{RPA}}{Q^\dag}{\mbox{RPA}}$ nor
$\mean{\mbox{RPA}}{Q}{\mbox{RPA}}$ can be different from zero.
Thus, we are left to calculate the expectation values
$\mean{\mbox{RPA}}{b^\dag_{p'} b_p}{\mbox{RPA}}$ and
$\mean{\mbox{RPA}}{b^\dag_{h'} b_h}{\mbox{RPA}}.$ The derivation
was first given in Ref.\cite{Rowe} and we will not repeat it here.
In spherical symmetry, the above expectation values are non zero
only if the states $p$ and $p'$, or $h$ and $h'$, have identical
quantum numbers except for their energies. The final form of the
result makes use of Eq.(\ref{eq6}). We just give here the
expressions necessary for calculating the RPA correlated density:
\begin{equation}
    \mean{\mbox{RPA}}{\sum_{m_p}b^\dag_{p'} b_p}{\mbox{RPA}}
    = \sum_{\nu Jh}\hat J^2 Y^{(\nu J)}_{p'h}Y^{(\nu J)}_{ph}~,
\label{eq7}
\end{equation}
\begin{equation}
    \mean{\mbox{RPA}}{\sum_{m_h}b^\dag_{h'} b_h}{\mbox{RPA}}
    = \delta_{hh'}\hat j_h^2-\sum_{\nu Jp}\hat J^2 Y^{(\nu J)}_{ph}Y^{(\nu J)}_{ph'}~.
\label{eq8}
\end{equation}
These results are formally identical with those of usual
non-relativistic RPA. In the case of RRPA there is a major
difference, however, as already pointed out in Subsection 2.1: the
states $p,p'$ can belong to the unoccupied Fermi states $\{\bar
p\}$ or to the Dirac sea states $\{\tilde p\}$.

In general, the corrections due to RPA correlations suffer from a
double counting of the second order corrections, and therefore one
should subtract out this redondant contribution. This double
counting problem was analyzed by Ellis\cite{Ellis1970} who showed
that it comes from the second order exchange diagrams of the RPA
series. In the present case, however, we are dealing with a
Hartree-type theory for the mean field, and the RPA series
contains only diagrams of the direct type (summation of the bubble
diagrams). Therefore, there is no need for evaluating double
counting corrections in the present model.

With the help of Eqs.(\ref{eq7},\ref{eq8}) and using the set of
single-particle wave functions of RMF we can build the one-body
density $\rho({\bo r})$ of the correlated ground state. The first
term on the r.h.s. of Eq.(\ref{eq8}) gives rise to the
uncorrelated RMF density $\rho^{(0)}({\bo r})$, all other terms in
$YY$ correspond to corrections due to long range correlations of
RPA type.

\section{CALCULATIONS \label{sect3}}
In this work the calculations are performed with the effective
Lagrangian DD-ME2 determined by Lalazissis et al.\cite{DDME2} for
the RMF-RRPA approach. This model gives a very good description of
ground state properties at the (uncorrelated) level, and also it
accounts very well for the collective nuclear excitations in the
framework of the RRPA\cite{Nicsic2002}. We limit ourselves to the
case of spherical symmetry and we study the effects of long range
correlations on the density distributions of the three magic
nuclei $^{16}$O, $^{40}$Ca and $^{48}$Ca.

First, the self-consistent mean field is calculated by solving in
coordinate space the RMF coupled equations for the nucleonic and
mesonic fields. The detailed method is described for instance in
Ref.\cite{Meng2006}. This gives the complete set of
single-particle states which will be necessary for the RRPA
calculations. This complete set consists of the occupied states in
the Fermi sea ($h$ states), the unoccupied states above the Fermi
sea (${\bar p}$ states) and the empty states in the Dirac sea
(${\tilde p}$ states). There are many ${\bar p}$ states which are
unbound and therefore, we use a box boundary condition to
discretize the continuum. The box radius $R$ is chosen to be 12 fm
for $^{16}$O and 15 fm for the Ca nuclei.

Next, the RRPA solutions are obtained by constructing the secular
matrix $\displaystyle\lb\begin{array}{cc}A & B \\ -B &
-A\end{array}\rb$ in the particle-hole basis and diagonalizing it.
The necessary formalism is explained in Ref.\cite{Nicsic2002}. In
this configuration space approach one has to adopt a truncation of
the single-particle spectrum. In our case, we choose to keep all
bound Dirac states whereas for the Fermi states we keep all states
below $E_{cut}$= 120 MeV. Once the secular matrix is diagonalized
we obtain the RRPA states $\nu$ characterized by their amplitudes
$X^{(\nu)}$ and $Y^{(\nu)}$. In the present work we include all
the RPA states with natural parity $\pi=(-1)^J$ and angular
momenta up to $J=8$ in $^{16}$O, $J=10$ in $^{40,48}$Ca.

One important property of RPA built on top of a self-consistent
mean field is that the center-of-mass spurious state must be one
of the $J^{\pi}=1^-$ RPA states, and this spurious state must be
at zero energy. In our calculations, we indeed obtain such a state
and its RPA energy is very close (less than 500 keV) to zero. It
is well separated from other $J^{\pi}=1^-$ states. We can thus
clearly identify the spurious state and remove it from the states
contributing to the expressions (\ref{eq7}-\ref{eq8}).

\section{RESULTS AND DISCUSSION \label{sect4}}

In this section we present and discuss the effects of RPA
correlations on the densities of the doubly closed shell (or
subshell) nuclei $^{16}{\rm O}$, $^{40}{\rm Ca}$ and $^{48}{\rm
Ca}$. According to Eqs.(\ref{eq7}-\ref{eq8}) the correlated
densities $\rho(\bo r)$ can be decomposed into:
\begin{eqnarray}
\rho({\bo r})&=&\rho^{(0)}({\bo r})+\delta\rho_{part.}({\bo r})+\delta\rho_{hole}({\bo r})\nonumber\\
&\equiv&\rho^{(0)}({\bo r})+\delta\rho({\bo r})~, \label{eq9}
\end{eqnarray}
where $\rho^{(0)}({\bo r})$ is the uncorrelated RMF density, and
\begin{eqnarray}
\delta\rho_{part.}({\bo r})&=&\sum_{pp'}\varphi_{p'}^+({\bo r})\varphi_{p}({\bo r})\sum_{\nu Jh}\hat J^2 Y^{(\nu J)}_{p'h}Y^{(\nu J)}_{ph}~,\nonumber\\
\delta\rho_{hole}({\bo r})&=&-\sum_{hh'}\varphi_{h'}^+({\bo
r})\varphi_{h}({\bo r})\sum_{\nu Jp}\hat J^2 Y^{(\nu
J)}_{ph}Y^{(\nu J)}_{ph'}~. \label{eq10}
\end{eqnarray}
Here, $\varphi_{p}({\bo r})$ and $\varphi_{h}({\bo r})$ are the
single-particle wave functions of the RMF.

\subsection{DENSITIES OF $^{16}{\rm O}$ \label{subsect4.1}}

\begin{figure}[bth]
\centerline{\includegraphics[width=8cm]{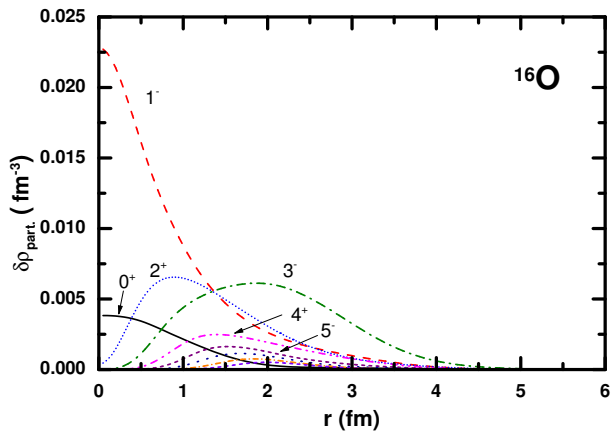}}
\caption{Contributions to $\delta\rho_{part.}$ of the $J^{\pi}$
multipoles from $0^+$ to $8^+$, in $^{16}$O. The unlabelled curves
correspond to $J^{\pi}=6^+,7^-,8^+$. \label{fig1}}
\end{figure}

\begin{figure}[bth]
\centerline{\includegraphics[width=8cm]{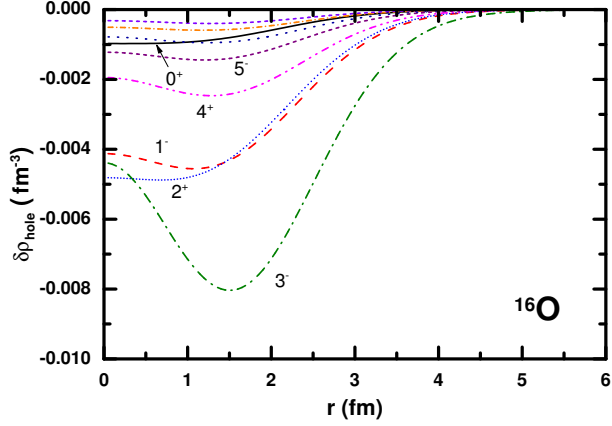}}
\caption{Same as Fig.1, for $\delta\rho_{hole}$ in $^{16}$O.
\label{fig2}}
\end{figure}

The respective contributions to $\delta\rho_{part.}$ and
$\delta\rho_{hole}$ of the various multipolarities $J^\pi$ of the
RPA states  are shown in Figs.\ref{fig1} and \ref{fig2}. The
calculations have been performed up to $J^{\pi}=8^+$. The
contributions to $\delta\rho_{part.}$ are generally positive
whereas those to $\delta\rho_{hole}$ are negative. The dominant
contributions come from the multipoles $1^-,2^+$ and $3^-$. This
is to be expected because the most collective RPA vibrations
belong to these multipoles. From Figs.\ref{fig1} and \ref{fig2} it
can be seen that the contributions decrease rapidly with
increasing $J$ and the convergence is reached at $J^{\pi}=8^+$.

\begin{figure}[bth]
\centerline{\includegraphics[width=8cm]{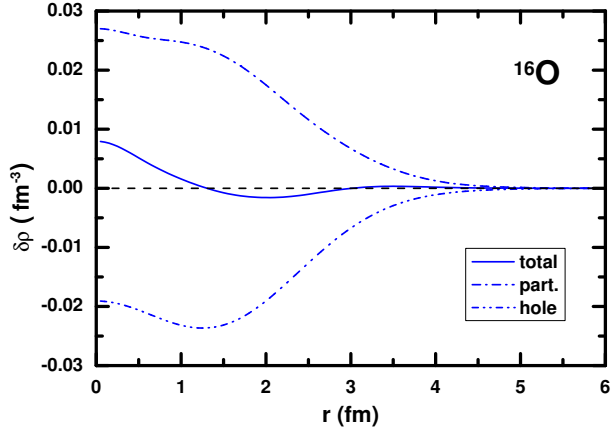}}
\caption{The summed corrections $\delta\rho_{part.}$,
$\delta\rho_{hole}$ and their total  $\delta\rho$ in $^{16}$O.
\label{fig3}}
\end{figure}

Summing over the partial contributions we obtain the corrections
$\delta\rho_{part.}$,  $\delta\rho_{hole}$ and $\delta\rho$ which
are displayed in Fig.\ref{fig3}. There are strong cancellations
between $\delta\rho_{part.}$ and $\delta\rho_{hole}$, and the net
result is small. We note that the analytical structure of
Eqs.(\ref{eq7}-\ref{eq8}) insures that the integral of
$\delta\rho$ over space is zero and hence, the total number of
neutrons and protons is not affected by the RPA corrections. This
is well verified in our numerical results.

\begin{figure}[bth]
\centerline{\includegraphics[width=8cm]{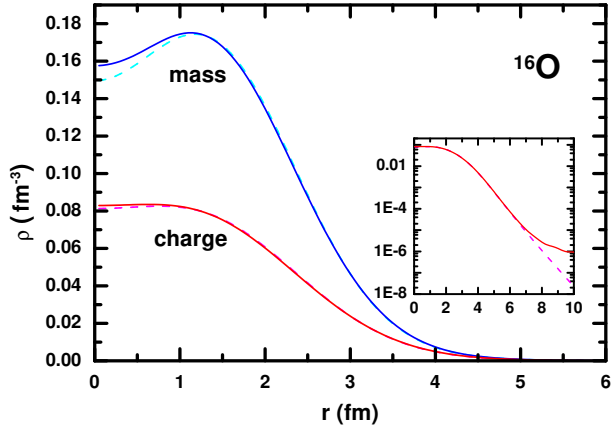}} \caption{The
uncorrelated (dashed lines) and correlated (solid lines) mass and
charge densities in $^{16}$O. In the inset are shown the charge
densities in log scale. \label{fig4}}
\end{figure}

In Fig.\ref{fig4} are shown the mass and charge densities, both in
the correlated (RPA) and uncorrelated (RMF) cases. For the charge
density the effects of the center-of-mass motion and of the finite
proton size are included by the usual folding
procedure\cite{Campi1972}. For the mass density $\delta\rho(r)$
the correlations tend to fill up the hole in the central region
while the surface becomes slightly steeper. The same tendency
appears in the charge density to a lesser extent. In the inset of
Fig.\ref{fig4} the charge density is shown in logarithmic scale in
order to display the differences in the tails in the asymptotic
region beyond 6 fm. It is seen that the correlations induce in the
outer region a tail which decreases less rapidly than the
uncorrelated density. To have a more global measure of the changes
caused by the RRPA correlations, the root mean square (rms) radii
of the densities are shown in Table \ref{Tab1}. The changes on
radii appear to be modest, of the order of 0.7\%-0.8\%.

\begin{table}[tbh]
    \caption{Mass and charge rms radii in $^{16}$O and $^{40}$Ca, calculated with
        uncorrelated RMF and correlated RPA densities.
        All units are in fm.}
    \vspace{.3cm}
    \begin{center}
    \begin{tabular}{lccc}\hline\hline
        Nucleus & &
        $\lc r^2_{\mbox{\scriptsize m}} \rc^{1/2}$ &
        $\lc r^2_{\mbox{\scriptsize ch}} \rc^{1/2}$ \\  \hline
        $^{16}$O  & RMF & $2.594$ & $2.729$ \\
                  & RMF+RPA & $2.614$ & $2.747$ \\ \hline
        $^{40}$Ca & RMF & $3.346$ & $3.466$ \\
                  & RMF+RPA & $3.385$ & $3.501$ \\
        \hline\hline
    \end{tabular}
    \label{Tab1}
    \end{center}
\end{table}

\subsection{DENSITIES OF $^{40}{\rm Ca}$ \label{subsect4.2}}

\begin{figure}[bth]
\centerline{\includegraphics[width=8cm]{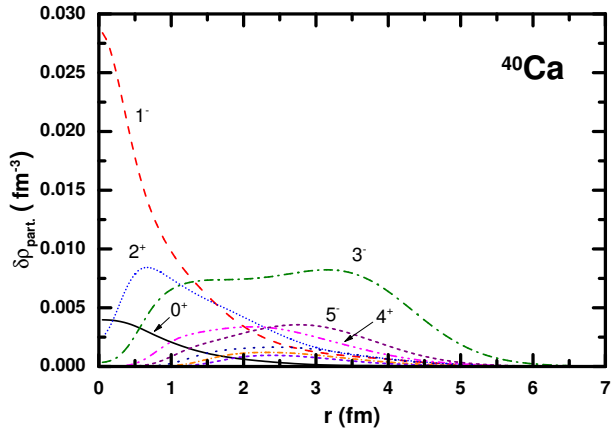}}
\caption{Contributions to $\delta\rho_{part.}$ of the $J^{\pi}$
multipoles from $0^+$ to $10^+$, in $^{40}$Ca. The unlabelled
curves correspond to $J^{\pi}=6^+,7^-,8^+,9^-,10^+$. \label{fig5}}
\end{figure}

\begin{figure}[bth]
\centerline{\includegraphics[width=8cm]{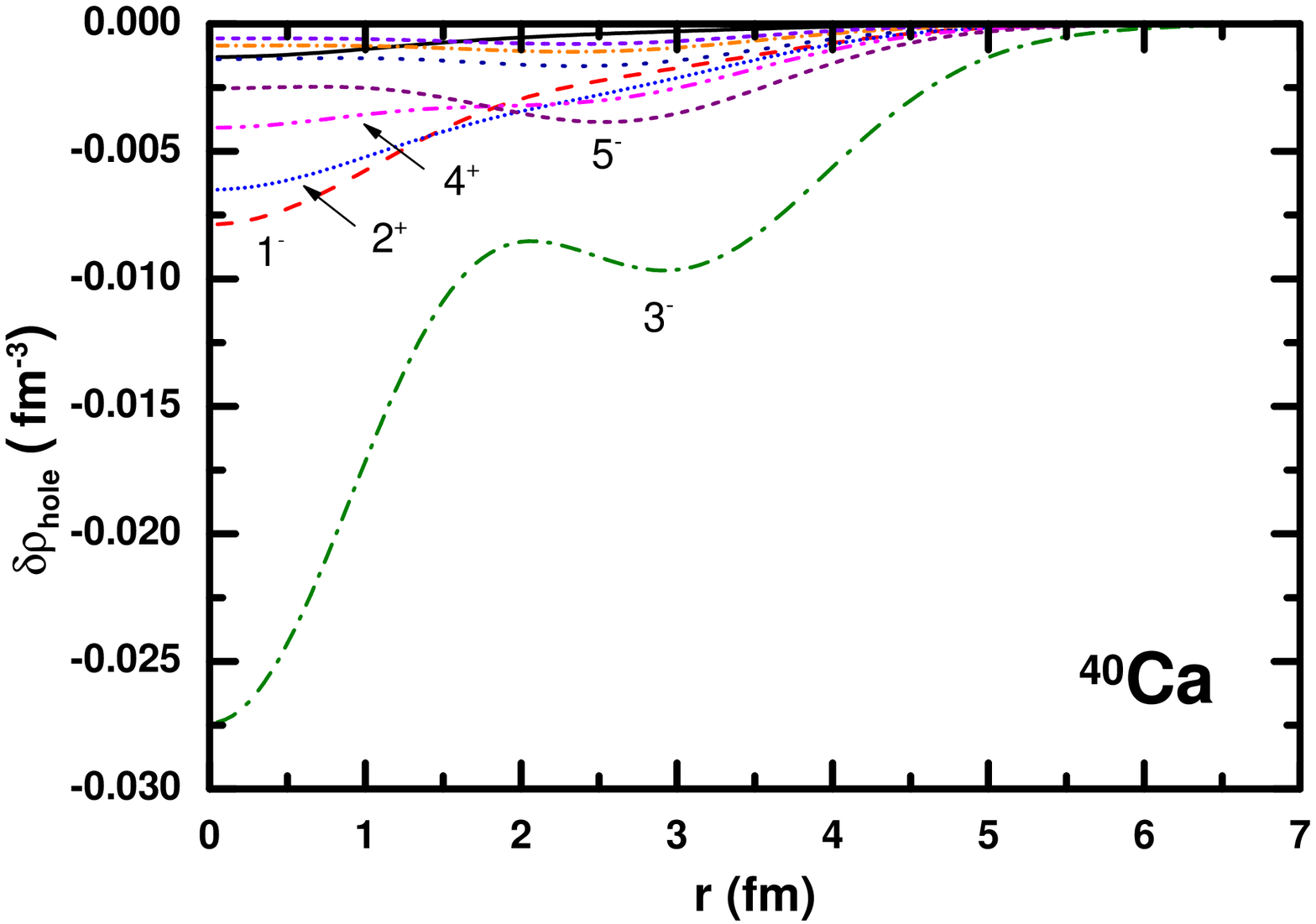}}
\caption{Same as Fig.5, for $\delta\rho_{hole}$ in $^{40}$Ca.
\label{fig6}}
\end{figure}

\begin{figure}[bth]
\centerline{\includegraphics[width=8cm]{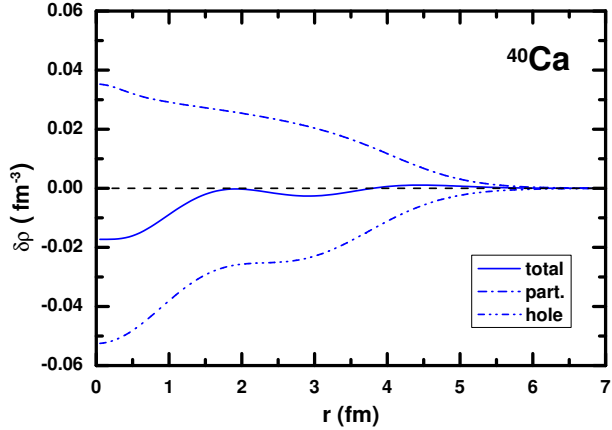}}
\caption{Same as Fig.3, for $^{40}$Ca. \label{fig7}}
\end{figure}

\begin{figure}[bth]
\centerline{\includegraphics[width=8cm]{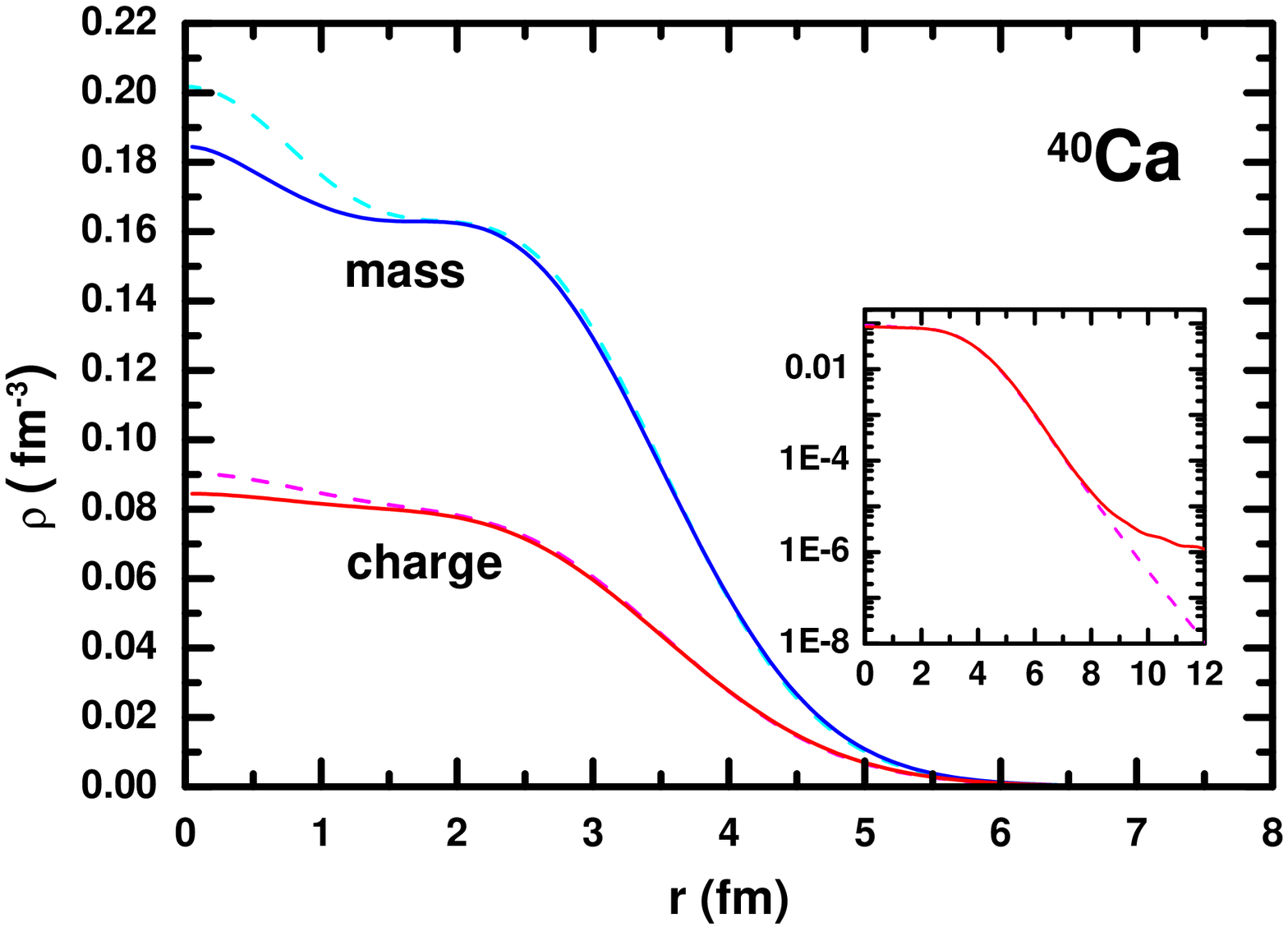}}
\caption{Same as Fig.4, for $^{40}$Ca. \label{fig8}}
\end{figure}

In Figs.\ref{fig5}-\ref{fig6} we present the respective
contributions to $\delta\rho_{part.}$ and $\delta\rho_{hole}$ of
the various multipolarities $J^\pi$ of the RPA states, in the
nucleus $^{40}{\rm Ca}$. Here again, the dominant contributions
come from the multipoles $1^-,2^+$ and $3^-$, and the convergence
is reached at $J^{\pi}=10^+$. It can be noted that the $1^-$ and
$2^+$ contributions are localized near the nuclear center whereas
the $3^-$ contributions extend over the nuclear volume. The summed
corrections $\delta\rho_{part.}$,  $\delta\rho_{hole}$ and
$\delta\rho$ are displayed in Fig.\ref{fig7}. The net effect on
$\delta\rho$ is still small because of the cancellations between
$\delta\rho_{part.}$ and $\delta\rho_{hole}$, but its shape is
different from that of the $^{16}O$ case: the correction is
negative from the center to about 3.5 fm, and positive afterwards
since the number of particles must be conserved. Thus, the change
on rms radii becomes slightly larger and of the order of 1\%, as
it can be seen in Table \ref{Tab1}. The mass density and charge
distribution (corrected for center of mass and finite proton size
effects as in the case of $^{16}$O) are shown in Fig.\ref{fig8}.

\subsection{DENSITIES OF $^{48}{\rm Ca}$ \label{subsect4.3}}

\begin{figure}[bth]
\centerline{\includegraphics[width=8cm]{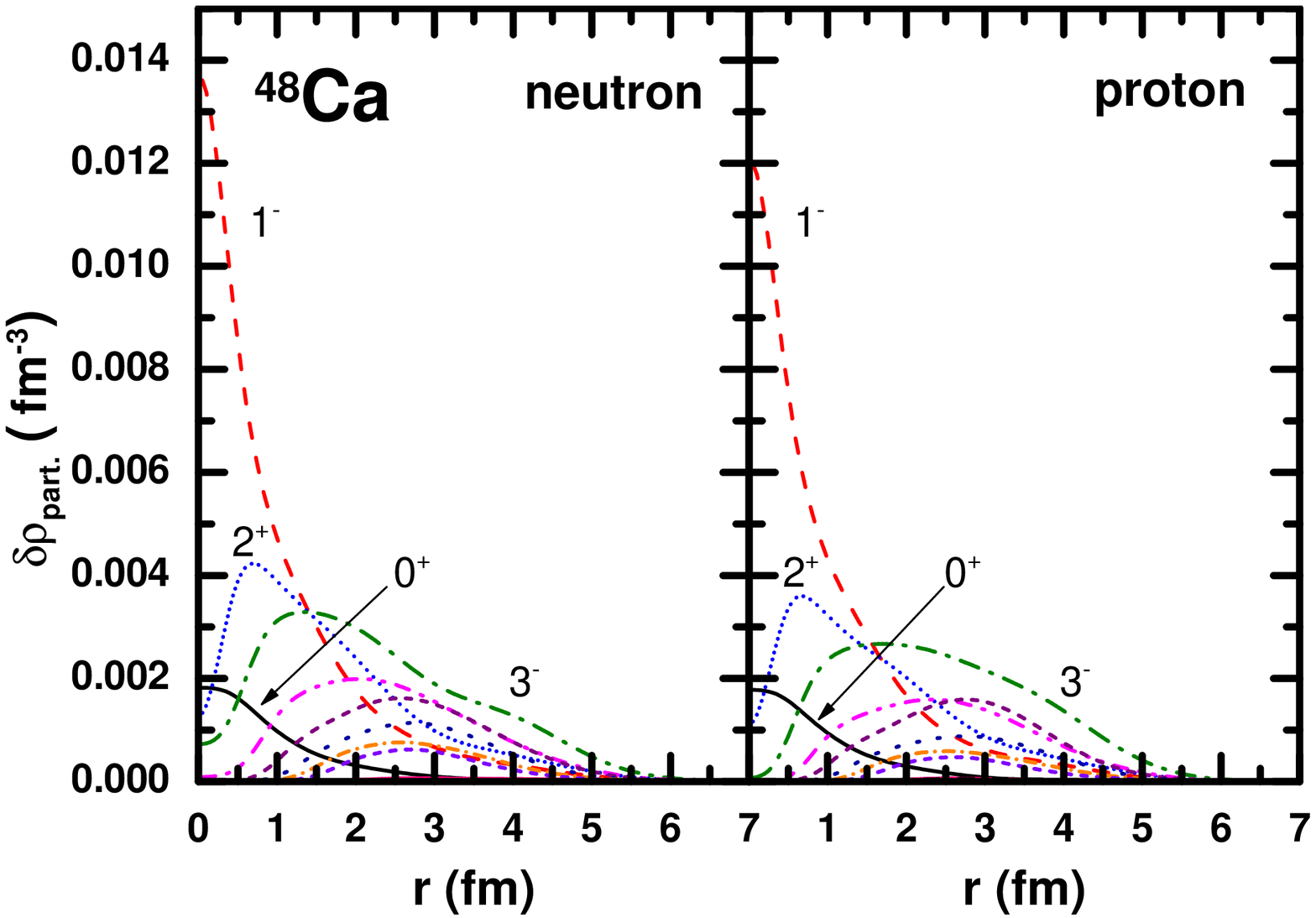}}
\caption{Contributions to $\delta\rho_{part.}^{(\nu)}$ (left
panel) and to $\delta\rho_{part.}^{(\pi)}$ (right panel) of the
$J^{\pi}$ multipoles from $0^+$ to $10^+$, in $^{48}$Ca. The
unlabelled curves correspond to
$J^{\pi}=4^+,5^-,6^+,7^-,8^+,9^-,10^+$. \label{fig9}}
\end{figure}

\begin{figure}[bth]
\centerline{\includegraphics[width=8cm]{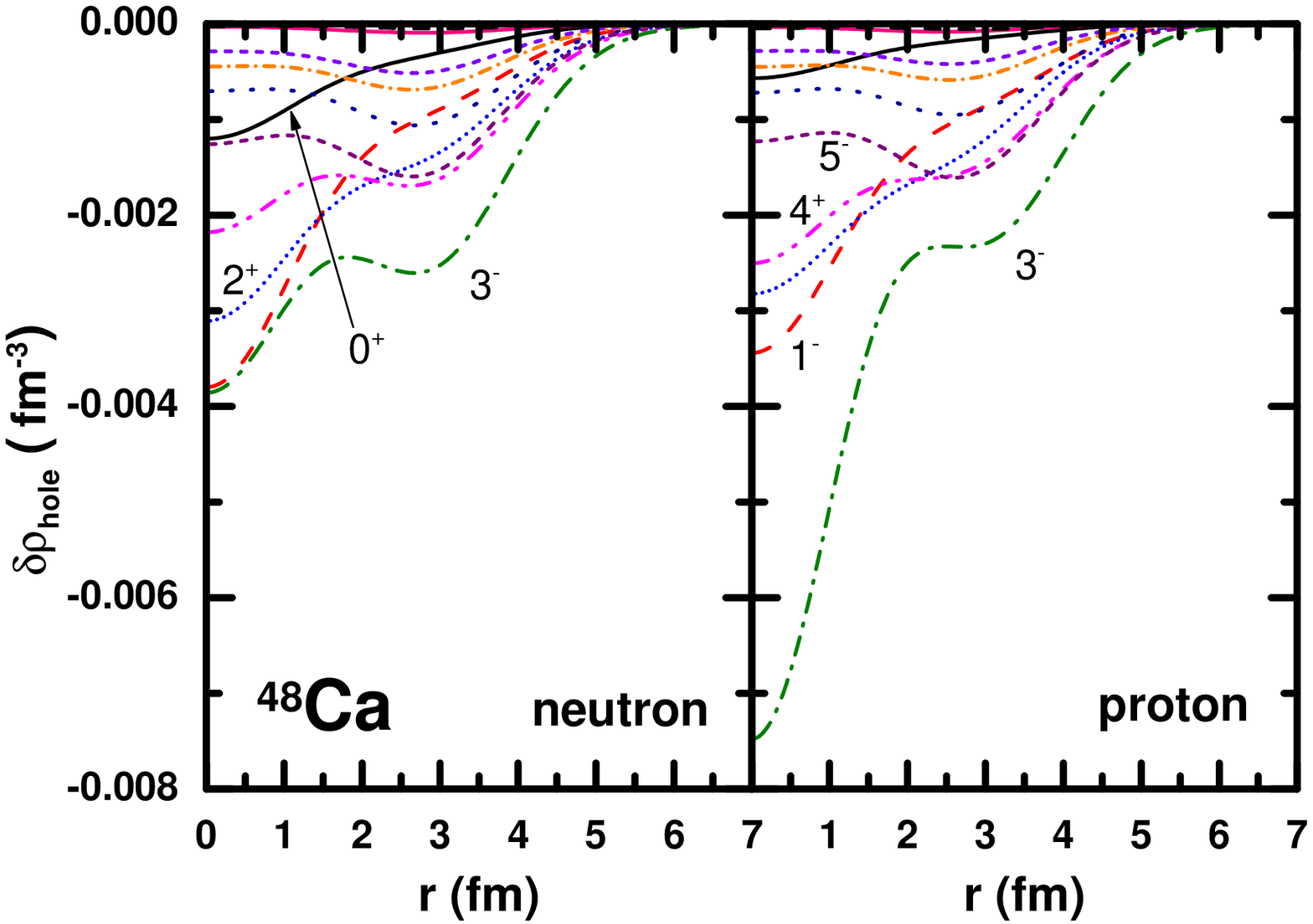}}
\caption{Same as Fig.{9}, for $\delta\rho_{hole}^{(\nu)}$ (left
panel) and  $\delta\rho_{hole}^{(\pi)}$ (right panel)
in $^{48}$Ca. \label{fig10}}
\end{figure}

For this nucleus with a sizable neutron excess we will examine
separately the neutron and proton distributions and show that the
long range correlations produce different effects on neutrons and
on protons. In Fig.\ref{fig9} are shown the partial contributions
of various multipoles to $\delta\rho_{part.}^{(\nu)}$ and
$\delta\rho_{part.}^{(\pi)}$. It can be seen that the different
multipoles contribute in a similar way to the neutron and proton
densities. Again, the dominant multipoles are those below $J=4$
and the convergence is reached before $J=10$. The situation is
slightly different for $\delta\rho_{hole}^{(\nu)}$ and
$\delta\rho_{hole}^{(\pi)}$ as one can see in Fig.\ref{fig10}. The
most important difference between neutron and proton densities is
seen in the contribution of the $J^\pi=3^-$ multipole. This can be
understood by looking at the expression (\ref{eq10}) for
$\delta\rho_{hole}$. In the case of neutrons the important
configurations $[(2s1/2,1d5/2,1d3/2)^{-1}(1f7/2)^1]_{3^-}$ are
missing while they are allowed for protons. This results in an
important contribution from $3^-$ states to the proton density
between 0 and 2 fm, as compared to the neutron density.

\begin{figure}[bth]
\centerline{\includegraphics[width=8cm]{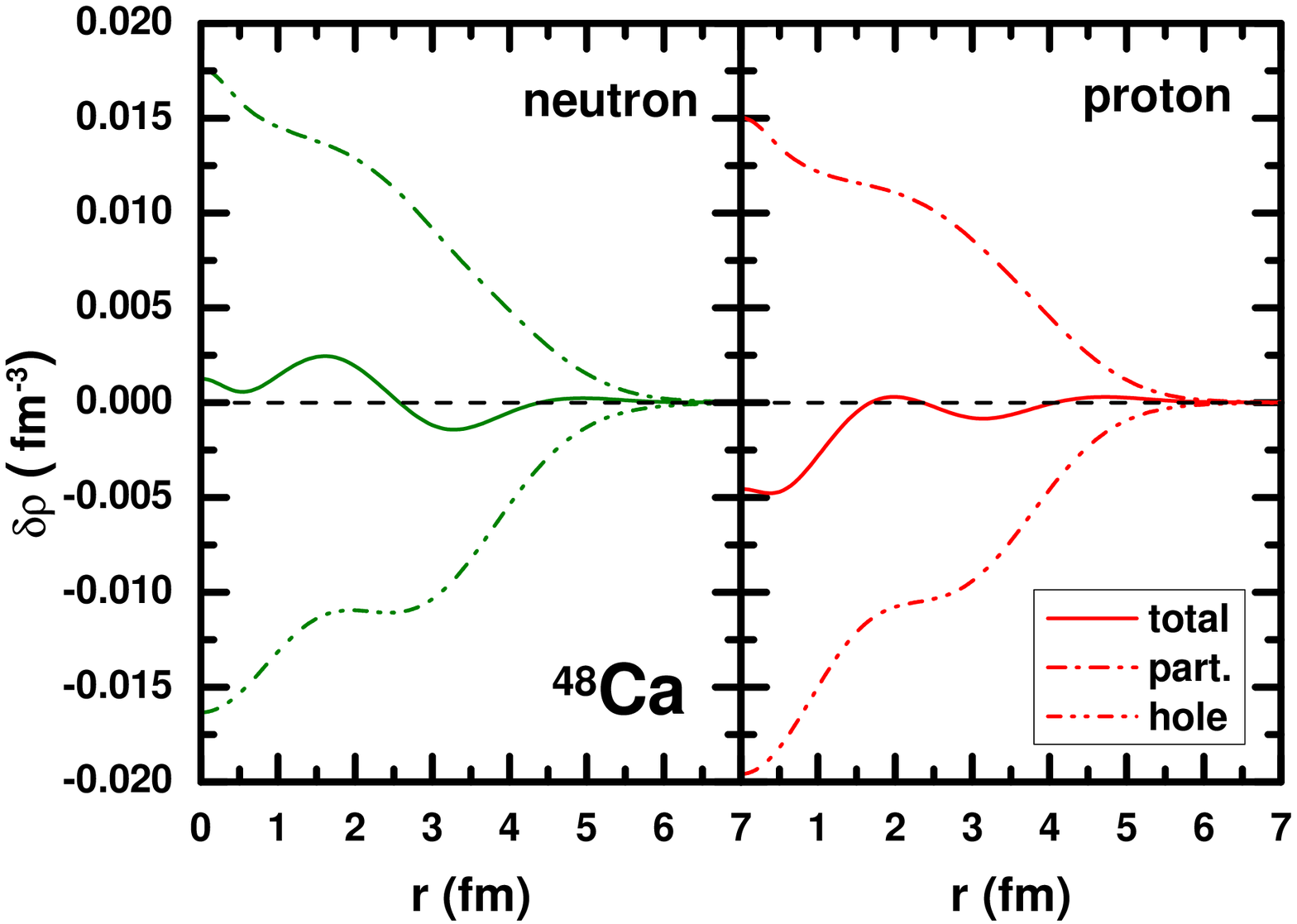}}
\caption{The corrections to neutron density (left panel) and
proton density (right panel) in $^{48}$Ca.  \label{fig11}}
\end{figure}

\begin{figure}[bth]
\centerline{\includegraphics[width=8cm]{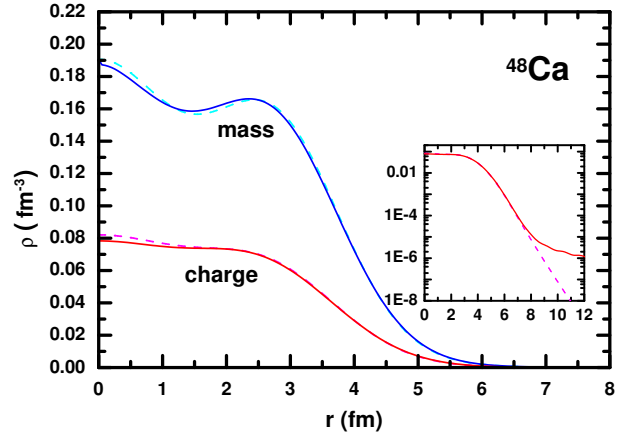}} \caption{
Same as Fig.4, for $^{48}$Ca.
\label{fig12}}
\end{figure}

Summing up the contributions of various multipoles we obtain the
density corrections $\delta\rho^{(\nu,\pi)}_{part.}$,
$\delta\rho^{(\nu,\pi)}_{hole}$ and their sum
$\delta\rho^{(\nu,\pi)}$. The results are shown in Fig.\ref{fig11}
for neutrons and protons separately. One can notice that the signs
of the corrections are opposite for neutrons and protons, in the
central region up to 2 fm. The corrections become negative between
2.5 to 4 fm, and finally they are positive in the outer part. In
Fig.\ref{fig12} we show the mass and charge densities of $^{48}$Ca
calculated with the uncorrelated (RMF) and correlated (RMF+RPA)
ground states. The general effect of correlations is to lower down
slightly the density near the center and to shift the
corresponding particles to the outer region. The exponent of the
exponential tail is changed somewhat beyond 8 fm. The values of
rms radii for uncorrelated and correlated densities are summarized
in Table \ref{Tab2}. The changes in neutron and mass radii are
relatively small (0.4\% and 0.6\%, respectively) whereas the
changes of proton density radius (0.9\%) and charge radius (0.8\%)
are more sizable.

\begin{table}[tbh]
    \caption{Neutron, proton, mass and charge rms radii in $^{48}$Ca, calculated with
uncorrelated RMF and correlated RPA densities.
        All units are in fm.}
    \vspace{.3cm}
    \begin{center}
    \begin{tabular}{lccccc}\hline\hline
        Nucleus & &
        $\lc r^2_{n} \rc^{1/2}$ &
        $\lc r^2_{p} \rc^{1/2}$ &
        $\lc r^2_{\mbox{\scriptsize m}} \rc^{1/2}$ &
        $\lc r^2_{\mbox{\scriptsize ch}} \rc^{1/2}$ \\  \hline
        $^{48}$Ca & RMF & $3.576$ & $3.389$ & $3.499$ & $3.482$\\
                  & RMF+RPA & $3.592$ & $3.420$ & $3.521$ & $3.511$\\
        \hline\hline
    \end{tabular}
    \label{Tab2}
    \end{center}
\end{table}

\begin{figure}[bth]
\centerline{\includegraphics[width=8cm]{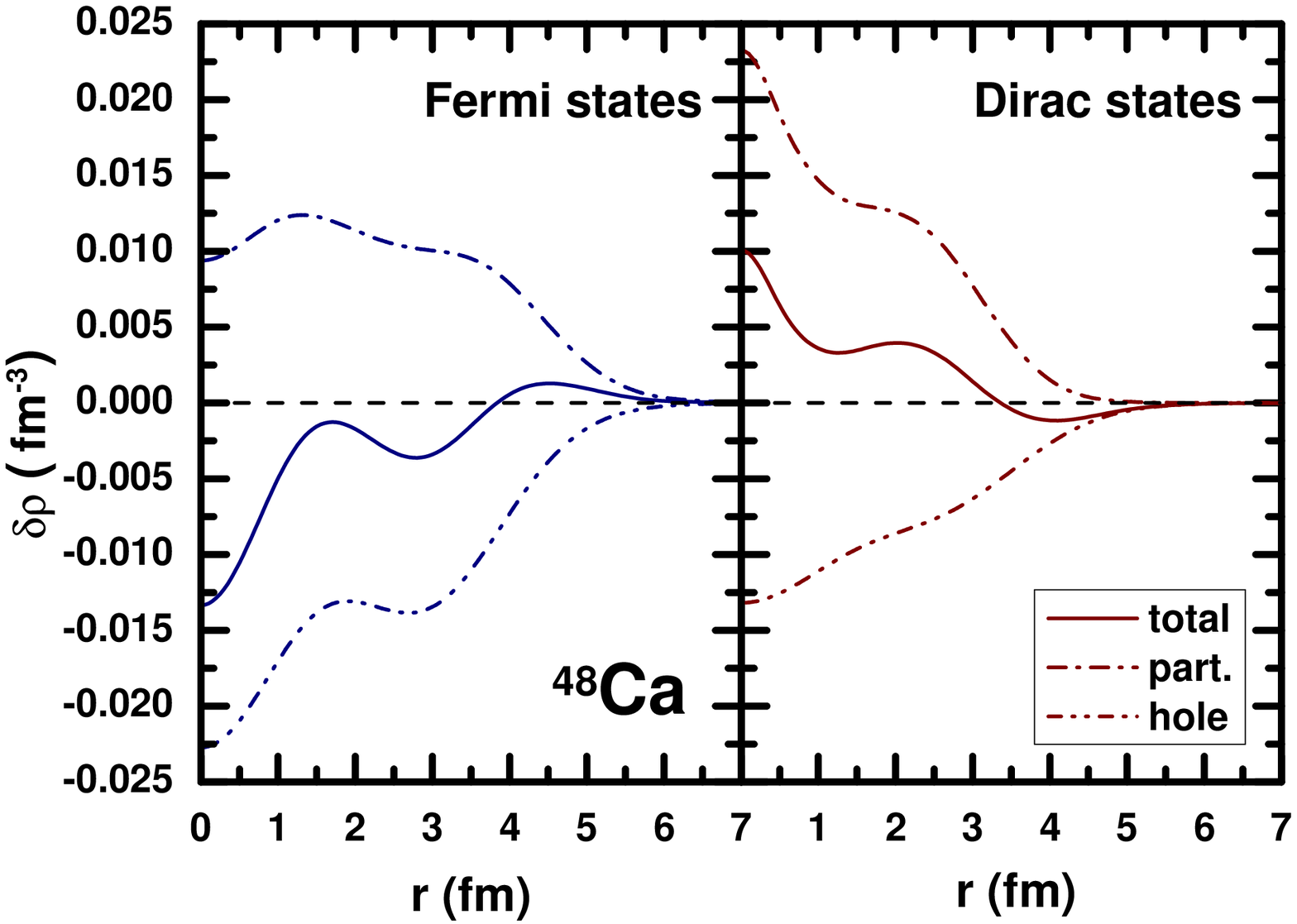}}
\caption{The contributions of Fermi states (left panel) and Dirac
states (right panel) to the mass densities $\delta\rho_{part.}$,
$\delta\rho_{hole}$ and $\delta\rho$ in $^{48}$Ca. \label{fig13}}
\end{figure}

Finally, it is interesting to consider the respective roles of the
Fermi and Dirac states in generating the corrections to the
densities, in the case of RRPA correlations. In Fig.\ref{fig13} we
show the separate contributions of the Fermi sector and Dirac
sector to the density corrections $\delta\rho_{part.}$,
$\delta\rho_{hole}$ and their sum $\delta\rho$. It can be seen
that the contributions of the Fermi and Dirac sectors to
$\delta\rho_{part.}$ are both positive whereas they are both
negative for  $\delta\rho_{hole}$. The magnitudes of Fermi and
Dirac corrections are comparable. Thus, the Fermi and Dirac
sectors play equally important roles in the density corrections.
The fact that the total correction $\delta\rho$ remains small is
due to the large cancellation between the positive
$\delta\rho_{part.}$ and the negative $\delta\rho_{hole}$.


\section{CONCLUSION \label{sect5}}
In this work we have calculated the effects of RPA correlations on
the ground state densities of some closed shell and closed
subshell nuclei. This has been done for the relativistic mean
field model and relativistic RPA theory, using a density-dependent
effective Lagrangian with sigma, omega and rho couplings. We have
stressed the peculiarity of the relativistic approach as compared
to the usual non-relativistic approach, with the important effects
coming from the negative energy (Dirac) sector. The presence of
the Dirac states is inherent to the relativistic approach because
one needs the completeness of the single-particle basis.

The long range correlation effects on densities are found to be
limited although not negligible. It turns out that the main
important multipolarities are the low $J$ modes, which is to be
expected because the important collective excitations correspond
to these low multipoles. The partial contributions to densities
decrease rapidly with increasing $J$, and it is sufficient to drop
the contributions beyond $J=8$ in $^{16}$O and $J=10$ in
$^{40,48}$Ca.

We have used in this work the DD-ME2 effective Lagrangian which
gives a good description of nuclear ground states in the RMF
approach, and it is working well for collective excitations of
electric type, i.e., transitions of natural parity $\pi=(-1)^J$
described by RRPA without exchange terms\cite{Nicsic2002}.
However, it is not necessarily suitable for unnatural parity
excitations, and this is the reason why we have not considered in
this work the effects of magnetic transitions such as spin and
spin-isospin modes. This must await future work where the
effective Lagrangian is treated in Hartree+Fock
approximation\cite{Long2006} and where the possibility to have
pion- and $\rho$ tensor-coupling would enable one to describe on
the same footing the electric and magnetic transitions.








\subsubsection*{Acknowledgements}
This work is partly supported by the French CNRS program PICS
(contract no. 3473), the European Community project Asia-Europe
Link in Nuclear Physics and Astrophysics CN/Asia-Link 008 (94791),
and the National Natural Science Foundation of China under Grant
Numbers 10435010 and 10221003.

\end{document}